\documentclass[twocolumn,showpacs,showkey,nofootinbib]{revtex4}
\usepackage{easyReview}
\usepackage{comment}
\usepackage{media9}
\usepackage{animate}
\usepackage{graphicx}
\usepackage{cancel}
\usepackage{soul}
\usepackage{color}

\usepackage{amsmath}
\usepackage{amssymb}
\usepackage{pgf}
\usepackage{subfigure}
\definecolor{OliveGreen}{rgb}{0,0.6,0}

\newcommand{\be}{\begin{eqnarray}}
	\newcommand{\bel}{\begin{equation}\label}
		\newcommand{\ee}{\end{equation}}
	
	\newcommand{\barl}{\begin{eqnarray}\label}
		\newcommand{\ear}{\end{eqnarray}}
	\newcommand{\non}{\nonumber}
	
\begin{document}
		\title{Microwave pulse transparency in Flux-qubit based superconducting quantum metamaterial}		
		\author{S. Galovic$^1$, Z. Ivi\'c$^{1,2}$, V. Nikoli\'c Z$^{1}$. Pr\v zulj$^1$, D. Chevizovich$^1$, 
		}
		\affiliation{$^{1}$ Laboratory for Theoretical and Condensed Matter Physics, "VIN\v CA" Institute of Nuclear Sciences, National Institute of the Republic of Serbia, University of Belgrade P.O.Box 522, 11001 Belgrade, Serbia,\\
			$^{2}$Institute of Theoretical and Computational Physics, Department of Physics, 
			The University of Crete, P.O. Box 2208, Helion, 71003, Greece \\}
		\date{\today}
		
\begin{abstract}
	We consider the propagation of a classical microwave pulse through a simple setup of a quantum metamaterial composed of a large number of three-Josephson-junction flux qubits. We find that population inversion and electromagnetic waves propagate together as two-component nonlinear waves, exhibiting distinct features depending on the initial preparation of the qubit subsystem and the strength of the "matter"-light interaction.
	Three different regimes are observed. In the limit of weak nonlinearity, when all qubits are initially prepared in either the clockwise or counterclockwise persistent current state, population inversion undergoes coherent Rabi-like oscillations, with a complete transfer between these two opposite states. As nonlinearity approaches unity, the transition dynamics lose their oscillatory nature, and the system rapidly becomes frozen in a state of zero population inversion, where each qubit is trapped in a superposition with equal probabilities of clockwise and counterclockwise polarity. In the overcritical regime, population inversion exhibits pulsating behavior, but without complete transfer. In the extreme coupling limit, population inversion undergoes small-amplitude oscillations around its initial value, while the pulse amplitude oscillates around zero, indicating pulse stopping.
\end{abstract}



\maketitle
\section{Introduction}
\label{sec1}
Significant progress in nanotechnology over the past three decades has enabled the fabrication of mesoscopic structures that exhibit quantum behavior \cite{qub0,qub01,qub8,qub1,qub2,qub3,qub4,qub5, qub6, lind, qub7, cqed1,cqed2,cqed3,cqed4,cqed5,qmm,qmm1,qmm11,qmm10,qmm2,asai15,asai18,qmm3,qmm4,qmm5, qmm6,qi1,qi2,qi3,qi4}. These structures are superconducting (SC) circuits interrupted by one or more Josephson junctions (JJs). At low temperatures, SC circuits enter the quantum regime, as confirmed by experiments demonstrating energy level quantization, macroscopic quantum tunneling, and quantum superposition of clockwise (CW) and counterclockwise (CCW) persistent current states \cite{qub1,qub2,qub3,qub4}, among other phenomena.

Josephson junctions play a crucial role in the design of SC quantum circuits, introducing nonlinearity and enabling the realization of various types of effective two-level systems—superconducting quantum bits (QBs) \cite{qub0,qub01,cqed2,cqed3}. Operating at sub-Kelvin temperatures, QBs can be integrated into larger systems—quantum metamaterials (QMMs)—capable of maintaining global quantum coherence long enough to serve as building blocks for quantum technological devices \cite{cqed1,cqed2,cqed3,cqed4,cqed5,qmm1,qmm11,qmm10,qmm2,qmm3,qmm4,qmm5,qmm6,qi1,qi2,qi3,qi4} particularly in quantum communication and information processing \cite{qi1,qi2,qi3,qi4}.

A key advantage of QMMs over natural atomic systems is the tunability of their energy-level separation, which can be adjusted at will through an appropriate choice of circuit parameters or external magnetic fields. This tunability allows their interaction with electromagnetic (EM) radiation to be controlled on demand, making them highly suitable for applications involving the manipulation of electromagnetic wave propagation \cite{cqed1,cqed2,cqed3,cqed4,cqed5}[12–16]. For instance, classical microwave (MW) pulses have been used to control and measure qubit states in SC circuit-based quantum information technologies, employing so-called quantum nondemolition (QND) measurements.
\\
Apart from the mainstream research aimed at realizing quantum computers—an endeavor that remains highly challenging—the interaction between the electromagnetic (EM) field and "artificial matter" can give rise to a variety of novel phenomena. Understanding these effects requires exploring regimes beyond the weak coupling limit. For instance, several exotic phenomena have been theoretically predicted, including a "breathing" photonic crystal, a "quantum Archimedean screw" \cite{qmm1,qmm2,qmm11}, lasing in the microwave range \cite{asai15}, and a new state of matter—the quasi-superradiant soliton phase \cite{asai18}.

Due to the nearly macroscopic sizes of meta-atoms, studying MW propagation in quantum metamaterials (QMMs) could enhance our understanding of phenomena at the boundary between the quantum and classical worlds. In this context, MW-induced Rabi oscillations and Ramsey fringes \cite{qub7}, as well as the generation of nonclassical photon states in SQUID arrays cite{qmm10}, have been explored. Additionally, quantum interference phenomena, such as electromagnetically induced transparency (EIT) \cite{eit1,eit2} and self-induced transparency (SIT) \cite{sit1,sit2,sit3,sit4} in multi - qubit systems, have been investigated as potential mechanisms for slowing down light and mitigating decoherence. In particular, these studies have examined how relaxation processes—especially inhomogeneous broadening—can be harnessed for this purpose.

In this theoretical work, we study the propagation of a classical MW pulse through a QMM composed of multiple three-Josephson-junction (3-JJ) flux qubits (FQBs). Initially, all qubits are prepared in a state of equal polarity—either clockwise  or counterclockwise — with respect to the persistent current in the FQB loop. As the incoming pulse propagates through the QMM, it induces a reversal of the persistent current polarity, forming a unified wave. The characteristics of these complex waves can be controlled by adjusting the qubit parameters, the initial state of the qubit system, and the pulse velocity.

Our analysis is performed under idealized conditions, neglecting relaxation effects arising from environmental coupling, as well as structural inhomogeneities in the QMM due to fabrication-induced variations in Josephson junction (JJ) sizes.

In Section 2, we introduce the setup of a QMM composed of multiple flux qubits (FQBs) embedded in a one-dimensional SC transmission line and propose the corresponding mathematical model. In Section 3, we derive the fundamental system of evolution equations for the ‘atom’-field variables and, within the adiabatic approximation, reduce it to a single nonlinear (Duffing) equation for population inversion. Section 4 presents the solutions of this equation and analyzes their dependence on the initial conditions, pulse velocity, and system parameters. Finally, in Section 5, we summarize the main results and conclusions, including a discussion on the potential realization of an equivalent device using different types of qubits.
\section{Setup and mathematical model}
\label{2}
\begin{figure}[h]
		\includegraphics[width=9cm]{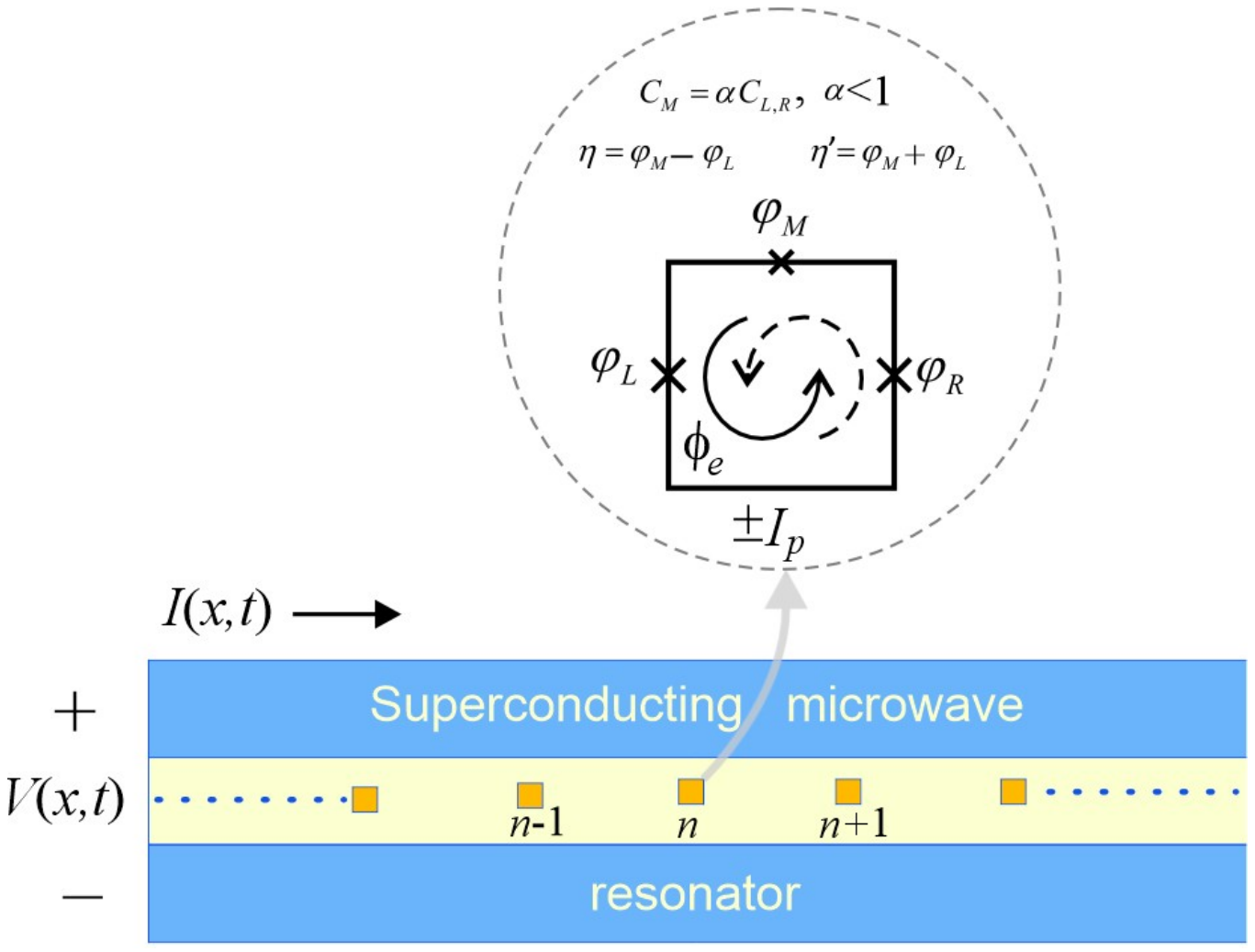}
	\caption{Schematic representation of a proposed setup: an infinite 1d array of superconducting flux qubits placed inside the infinite coplanar microwave resonator. Inset figure: a simplified scheme of 3-JJ flux qubit. Two side JJs are identical, while the central one has slightly less capacitance. $I_P$, $\Phi_e$, $\eta=\varphi_M-\varphi_L$ and $\eta'=\varphi_M+\varphi_L$ are persistent current in FQB loop, external flux threading the FQB loop and relative and total phase, respectively. }\label{setup}
\end{figure}

We consider the QMM that consists of an infinite coplanar two-stripe superconducting transmission line (TL) filled by an ensemble of $N\gg 1$ periodically arranged flux qubits  -  Figure (\ref{setup}). The inter-qubit distance ($l$) highly exceeds qubit size. Which, therefore, may be treated as point-like objects. The type of flux qubits is not essential For the present consideration. However, for the possible practical implementation of the proposed setup, being less susceptible to decoherence compared with the other types of FQBs, 3-JJ flux qubits are particularly suitable. At the same time, their quantum states can be manipulated via the external magnetic flux coming from the DC-current line, for example. In the concluding section, we will briefly discuss the realization of a similar setup employing the different types of qubits.
\subsection{Mathematical model: flux qubit}
Superconducting flux qubits are tiny, micrometer-sized, superconducting rings interrupted by one or more JJs. When an external magnetic flux is applied clockwise (CW) or counterclockwise (CCW), persistent current ($I_p$) will flow continuously through the ring. When the magnetic flux, enclosed in the ring, is comparable to half-integer values of SC flux quanta $\Phi_0$, potential energy versus Josephson's phase attains the form of asymmetric double-well potential \cite{fqb1,fqb2}. The two lowest levels corresponding to opposite circulating persistent currents are well separated from the excited states so that FQB is effectively a two-level quantum mechanical system whose Hilbert space is restricted to the basis $|\circlearrowright \rangle $, $|\circlearrowleft\rangle$ corresponding to CW or CCW polarity of persistent current. These states are eigenstates of the operator of the current through the flux qubit loop: $\hat{I}_{qb}=I_P \left(|\circlearrowright\rangle\langle \circlearrowright|-|\circlearrowleft\rangle\langle\circlearrowleft|\right))$. The persistent current magnitude ($I_P$) ranges approximately from one-half of the critical current in single JJ flux qubits to that of the critical current of the smallest of three junctions in the case of persistent current qubits \cite{lind}.
The Hamiltonian single FQB is given as \cite{fqb1,fqb2}:
\bel{sqb} H_{fq} =-\frac{\hbar\Delta}{2} \tau^z -\frac{\hbar h}{2} \tau^x. \ee
Here $h$ is the tunneling frequency between the wells, $\hbar\Delta = 2I_P (\Phi_e-\Phi_0 /2)$ is the energy bias due to external magnetic flux while $\tau$ -s are the Pauli matrices in persistent current basis:
\begin{eqnarray}
	\non\tau^x&=&|\circlearrowright\rangle\langle\circlearrowleft|+|\circlearrowleft\rangle\langle \circlearrowleft|,\\
	\tau_y&=&i(|\circlearrowright\rangle\langle\circlearrowleft|-|\circlearrowleft\rangle\langle \circlearrowleft|),\\
	\non \tau^z&=&|\circlearrowright\rangle\langle \circlearrowright|-|\circlearrowleft\rangle\langle\circlearrowleft|.
\end{eqnarray}
Apparently, CW and CCW states are eigenstates of $\tau^z$ with eigenvalues $\pm 1$:
 $\tau^z |\circlearrowright\rangle=  |\circlearrowright\rangle$ and $\tau^z |\circlearrowleft\rangle= - |\circlearrowleft\rangle $, while $\tau^x$ causes "spin" reversal  $\tau^x|\circlearrowleft\rangle = |\circlearrowright\rangle$ and  $\tau^x|\circlearrowright\rangle = |\circlearrowleft\rangle$. 
A multi-qubit subsystem Hamiltonian has the form:
\bel{hqb}
H_{fq} = -\frac{\hbar}{2}\left[ \sum_n \left( \Delta\tau^z_n + h \tau^x_n\right)\right] , 
\ee
\subsection{Mathematical model: Transmission line}
The coupling of qubits with the EM field in the transmission line is due to the mutual inductance of the current circulating through the qubit loop ($\hat{I}_{qb}=I_P\tau_z$) and the current in the transmission line. We now employ the lumped-element model circuit model (LECM) to find the interaction Hamiltonian \cite{pozar}. In such an approach, the TL is usually represented schematically as a two-wire line divided into an infinite number of parts of length ($\Delta x$). Each of these pieces represents the fictitious LC circuit whose elements are specified through per-unit-length inductance and capacitance  $L$ and $C$, as follows: $\Delta L=L\Delta x$ and $\Delta C=C\Delta x$. Such an approach enables one to employ the standard theory of electric circuits \cite{pozar} and to, in the final instance, i.e., after passing back to the continuum ($\Delta x\rightarrow 0$), demonstrate wave-like propagation of voltage and electric current through the TL. Specifically, system dynamics is described by a set of \textit{telegrapher's equations} which straightforwardly lead to classical wave equations for currents and voltages. The effective model is Hamiltonian, and currents and voltages play the role of canonical variables \cite{roth,pozar}.

\begin{figure}[h]
\centering
	\includegraphics[width=7cm]{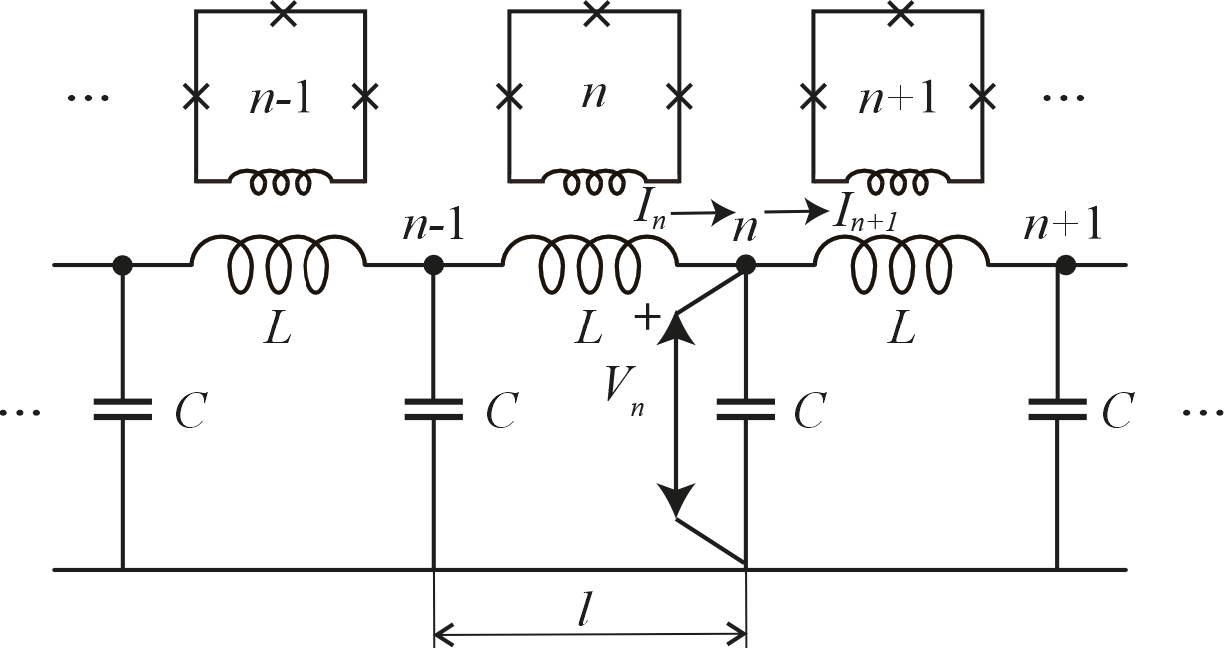}
\caption{Discretized representation of proposed setup.}\label{discr}
\end{figure}

An analogous procedure may be applied here. Specifically, we formally discretize TL and represent it as an infinite chain of fictitious LC circuits of length equal to inter-qubit distance so that $\Delta x \approx l$ - Fig.(\ref{discr}). Transmission line dynamics is described by a set of discrete variables: electric currents ($I_n(t)$) and voltages ($V_n(t)$). Also, to each node $n$ is assigned a flux and charge variables: $\varphi_n=\int_{-\infty}^{t} dt'V_n(t')$ and $\mathcal{Q}_n=C(\partial \varphi_n/\partial t)\equiv CV_n$.

We now formulate the Hamiltonian of TL. To this end, we apply Kirchhof's laws for voltage and currents at the node $n$ to obtain a system of evolution equations for voltages and currents. In this way, we got the system of difference-differential equations - discrete telegrapher equations: 
\barl{kirch}
\non \frac{\partial I_n(t)}{\partial t}=-\frac{1}{L}(V_{n+1}(t)-V_{n}(t)),\\
\frac{\partial V_{n+1}}{\partial t}=-\frac{1}{C} (I_{n+1}(t)-I_{n}(t)).
\ear
Combining the first equation with the definition of node flux, which implies ${\dot\varphi_n(t)}=V_n(t)$, we obtain:
\bel{curr}
I_n(t)=-\frac{1}{L}(\varphi_{n+1}-\varphi_n).
\ee
Now combining, the last result with the equations (\ref{kirch}) and accounting for the above definitions, we deduce the evolution equation for node flux:
\barl{fluxeq}
\ddot \varphi_n(t)-\frac{1}{LC}(\varphi_{n+1}+\varphi_{n-1}-2\varphi_n)=0.
\ear 
It is easy to show that the last equation follows straightforwardly from the following Hamiltonian 
\bel{htl}
H_{TL}=\frac{1}{2}\sum_n 
\left[\frac{\mathcal{Q}^2_n}{C}+\frac{1}{L}(\varphi_{n+1}-\varphi_n)^2\right]. 
\ee
Here, flux ($\varphi_n$) and charge ($Q_n$) play the role of generalized "coordinate" and "momentum":
\barl{he}
\frac{d \mathcal{Q}_n}{dt}=-\frac{\partial\mathcal{H}}{\partial \varphi_n}, &\;& \frac{d \varphi_n}{dt}=\frac{\partial\mathcal{H}}{\partial \mathcal{Q}_n}.
\ear
\subsection{Mathematical model: Interaction}
To derive the Hamiltonian of an interaction, we first consider the single qubit placed at $n$-th node. Its interaction with the resonator is $\sim I^{qb}_nI_n$ where the $I^{qb}_{n}=I_{P}\tau^z_n$ is the persistent current in the $n$-th qubit, while $I_n$ is the current in the segment of resonator closest to that qubit. Introducing the mutual inductance ($M$) and employing equation (\ref{curr}) we found $H_i(n)=MI_{P}\tau^z_n I_n$. 
by summing over all nodes, we get the total Hamiltonian of the qubit subsystem
\barl{int} H_i=\frac{g}{2}\sum_n \tau^z_n(\varphi_{n+1}-\varphi_n).
\ear
The coupling parameter reads: $g=2MI_p/L$. Factor 2 is included here for the mathematical convenience. 

Concluding this chapter, we recall that the discretization procedure outlined here is an auxiliary (intermediate) step, while the transition back to the continuum is understood as the final step of this procedure. In the present case, where we deal with an intrinsically discrete qubit subsystem embedded in a continuous medium, transition to continuum holds provided that the wavelength of EM radiation ($\lambda$) highly exceeds inter-qubit separation $\lambda \gg l$. Moreover, all dynamical variables should slowly vary over $l$. Otherwise, accounting for the discreteness of the qubit subsystem is necessary. For example, when qubit size is comparable with interqubit distance, they cannot be treated as point-like objects. In this case,  interaction Hamiltonian has the form:
\barl{intmix} H_i=\frac{g}{2}\sum_m \tau^z_m\int\frac{d x}{l}\frac{\partial \varphi(x,t)}{\partial x} \delta(x-ml).\ear 
The $m$ corresponds to the qubit position, and the sum runs over whole qubits.

\section{Dynamical equations}\label{iii}

We now derive the evolution equations describing the propagation of the classical EM wave through the above-proposed "device." To simplify practical calculations we here employ the time-dependent variational principle \cite{roth} and derive the evolution equations requiring 
the stationarity ($\delta A=0$) of action functional $A=\int_{t_1}^{t_2}dt \mathcal{L}(\Psi,\Psi^*; t)$ where $\mathcal{L}(\Psi,\Psi^*; t)=
\frac{i\hbar}{2}(\langle \Psi(t)|\dot\Psi(t)\rangle-\langle\dot\Psi(t)|\Psi(t)\rangle)-\langle\Psi|H_T|\Psi\rangle$ 
denotes the system Lagrangian in which the last term plays the role of classical Hamiltonian. Here $H_T=H_{q}+H_i+H_{TL}$ is the total Hamiltonian.
The vector of state we took in the form: \bel{vec}|\Psi(t)\rangle=\sum_n\left( \Psi^{+}_n(t)|\circlearrowright\rangle+\Psi^{-}_n(t)|\circlearrowleft\rangle \right),\ee
corresponding to qubit in a superposition of CW and CCW polarized persistent current state. Complex functions $\Psi^{\pm}_n(t)$  are corresponding probability densities. They, at each node satisfy the following relation $|\Psi^{+}_n|^2+|\Psi^{-}_n|^2=1$. Accordingly the vector of state (\ref{vec}) is normalized as:
\bel{norm}
\langle \Psi|\Psi\rangle=\sum_n |\Psi^{+}_n|^2+|\Psi^{-}_n|^2=\sum_n 1\equiv N.
\ee
Transition to functions normalized unity is easy through simple transformation $\Psi^{\pm}\rightarrow \frac{1}{\sqrt{N}}\Psi^{\pm}$. Nevertheless, it is not essential for further analysis since it would result in the same system of evolution equations.

Substitution of (\ref{vec}) into action functional followed with the requirement $\delta A=0$, would result in the system of difference-differential evolution equations for qubit ($\Psi^{\pm}_n(t)$) and EM field ($\varphi_n(t), \; Q_n(t)$) variables. Consistent with the discussion at the end of the preceding chapter, we will carry on our study in continuum approximation. In such a way, all discrete functions of the form $f_{n\pm 1}$ became $f(x\pm l)$ where $l$ is considered small compared to the spatial extension of the function itself. In this way passing to continuum via $f_{n\pm 1} \longrightarrow f(x\pm l)\approx f(x)\pm l \frac{\partial f}{\partial x}+\frac{l^2}{2!}\frac{\partial^2 f }{\partial x^2}+...$ is satisfactory. 

In continuum approximation, all dynamical variables satisfy Hamilton's equations:

\barl{he}
\non i\hbar\frac{d \Psi^{\pm}}{ dt}&=&\frac{\delta\mathcal{H}}{\delta \Psi^{* \pm}},\\
\frac{d \mathcal{Q}}{dt}=-\frac{\delta\mathcal{H}}{\delta \varphi}, &\;& \frac{d \varphi}{dt}=\frac{\delta\mathcal{H}}{\delta \mathcal{Q}},
\ear
where 
\barl{hamf}
\non && \mathcal{H}=N \int_{-\infty}^{\infty}\frac{dx}{l}\left[\dfrac{Q^2}{2C}+\frac{1}{2L}\left( \dfrac{\partial \varphi}{\partial x}\right) ^2 \right]-\\
&&  \dfrac{g\hbar l}{2} \int_{-\infty}^{\infty}\frac{dx}{l}(|\Psi^{+}|^2-|\Psi^{-}|^2)\left(\dfrac{\partial \varphi}{\partial x}\right) +\\
\non &&\int_{-\infty}^{\infty} \frac{dx}{l}\left[ \dfrac{\hbar\Delta}{2}(|\Psi^{+}|^2-|\Psi^{-}|^2)+\dfrac{\hbar h}{2}(\Psi^{*+}\Psi^{-}+c.c.)\right],
\ear
plays the role of classical Hamilton's function. At the same time, the variation of any of these variables is specified 
\barl{var}\frac{\delta \mathcal{H}}{\delta B}=\frac{\partial \mathcal{H}}{\partial B}-\dfrac{d}{d t}\frac{\partial \mathcal{H}}{\partial \dot B}-\frac{d}{dx}\frac{\partial\mathcal{H}}{\partial B'}. 
\ear
Here, we associate $B\equiv B(x,t)$ with any dynamical variables, while primes and dots correspond to derivation over $x$ and $t$, respectively.

Employing Hamilton's equations (\ref{he}) we obtain the following set of coupled 'Maxwell-Schr\"odinger equations':
\barl{maxw}
\non
i\dot\Psi^{+}&=&-\frac{\Delta}{2}\Psi^{+}-\frac{h}{2}\Psi^{-}+\frac{gl}{2}\frac{\partial \varphi }{\partial x}\Psi^{+},\\
i \dot\Psi^{-}&=&\frac{\Delta}{2}\Psi^{-}-\frac{h}{2}\Psi^{+}-\frac{gl}{2}\frac{\partial \varphi  }{\partial x}\Psi^{-},\\
\non \ddot\varphi&=& s^2 \frac{\partial^2 \varphi}{\partial x^2} + \frac{\hbar g l}{2C N} \frac{\partial }{\partial x}\left( |\Psi^+|^2-|\Psi^-|^2\right),
\ear
where $s=l/\sqrt{LC}$ is the speed of EM radiation in the empty QMM.

To simplify further calculation we now introduce Bloch (Stockes) variables corresponding to normalized per particle expectation values of Pauli matrices in state $|\Psi(x,t)\rangle$: $ P(x,t)=\langle \Psi| \tau^z|\Psi\rangle/\langle \Psi|\Psi\rangle \equiv (|\Psi^{+}(x,t)|^2-|\Psi^{-}(x,t)|^2)/N$, $R(x,t)=\langle \Psi| \tau_{n,x}|\Psi\rangle/\langle \Psi|\Psi\rangle\equiv (\Psi^{*+}(x,t)\Psi^{-}(x,t)+c.c.)/N \; \mathrm{and}\; Q(x,t)=i(\Psi^{*+}(x,t)\Psi^{-}(x,t)-c.c)/N$.

Combining the first two equations in (\ref{maxw}), we find the following system of equations:
\barl{nlbloch}
\non &&\dot R(x,t)=\Delta Q(x,t)-gl \frac{\partial \varphi(x,t)}{\partial x}  Q(x,t),\\
\non &&\dot Q(x,t)=-\Delta R(x,t)+gl \frac{\partial \varphi(x,t)}{\partial x}R(x,t)+hP(x,t),\\
&&\dot P(x,t)=-hQ(x,t),\\
\non &&\ddot\varphi(x,t)-s^2 \frac{\partial^2 \varphi(x,t)}{\partial x^2}= \frac{\hbar g l}{2C} \frac{\partial P(x,t)}{\partial x},
\ear
In what follows we will consider self-consistent solutions of the above system assuming steady state propagation of joined MW pulse and \emph{population inversion} $\left(P(x,t)\right))$ representing the transition probability between the two states of qubit corresponding to opposite polarity of the persistent current. In this picture the complex entity, MW pulse + population inversion, propagates trough the QMM in the form of a traveling wave depending on $t$ and $x$ only through the single variable - the time in moving frame $t \longrightarrow\tau=t-x/v$\footnote{Alternatively, one may introduce retarded coordinate in moving frame $\xi=x\pm vt$. These variables are connected as: $\tau=\mp \xi/v$.}. Transition to new variable yields: 
\barl{nlblochtau}
\label{nlblochtau}
\non && R_{\tau}(\tau)=\Delta Q(\tau)+\frac{gl}{v} \frac{\partial \varphi(\tau)}{\partial \tau}  Q(\tau),\\
\non && Q_{\tau}(\tau)=-\Delta R(\tau)-\frac{gl}{v} \frac{\partial \varphi(\tau)}{\partial \tau}R(\tau)+hP(\tau),\\
&& P_{\tau}(\tau)=-hQ(\tau),\\
\non &&\varphi_{\tau\tau}(\tau) =\frac{\hbar glv}{2 C s^2 (1-\frac{v^2}{s^2} )}P_{\tau}(\tau),
\ear
The last equation in the above system may be integrated in the adiabatic approximation which holds provided that the aforementioned steady state propagation condition is satisfied. Under these conditions the first integral of the last equation in system (\ref{nlblochtau}) reads:
\bel{I1}
\varphi_{\tau}=\frac{\hbar glv}{2Cs^2(1-v^2/s^2)}\left(P_0-P(\tau)\right)
\ee 
Here we kept only a particular solution, while the homogeneous part, responsible for fluctuations (\cite{ahmad,kaplan}) was disregarded. Such an approach has been a standard ansatz in self consistent treatment of phenomena where one encounters joined propagation of two different excitation. Typical example is self induced transparency (\cite{ahmad,kaplan}) in optics where light pulse is accompanied with population inversion of the ensemble of two-level atoms. Similar examples are studies of magneto-elastic \cite{magsol} waves and polaron mediated energy and charge transfer in molecular crystals \cite{polaron}, to mention just a few.

To determine integration constant, we assumed that initially ($\tau=\tau_0$) EM pulse entered the system and therefore has maximum value: $\varphi(\tau_0)\equiv \varphi_0$ and  ($\varphi_{\tau}(\tau_0)\equiv 0$). Also, we took  
that all qubits are prepared in CW or CCW state so that $|\Psi^{+}(\tau_0)|^2-|\Psi^{-}(\tau_0)|^2 = P(\tau_0)\equiv\pm 1$. In this way, we obtain a reduced system of three nonlinear Bloch equations:
\barl{nlb}
\non R_{\tau}=\delta Q-G(v)PQ,\\
Q_{\tau}=-\delta R+G(v)PR+hP,\\
\non P_{\tau}=-hQ
\ear 
Here $G(v)=\frac{\hbar g^2 l^2}{2s^2C(1-v^2/s^2)}$ is the nonlinearity parameter, while $\delta=\Delta +GP_0$ plays the role of the effective bias through the FQB loop modified by the contribution coming from the MW pulse.

By elimination of $R$ and $Q$ from the system (\ref{nlb}) we easily obtain a single nonlinear differential equation for population inversion. Thus, combining the third and the first equations, we obtain 
\bel{r-p}
R=\frac{\delta}{h}(P_0-P)-\frac{G(v)}{2 h}(P^2_0-P^2). 
\ee
Substitution the last result into a second equation of the system (\ref{nlb}), then employing the third equation from the system (\ref{nlb}), we eliminate $Q_{\tau}$ and obtain:
\barl{nlnd}
\non P_{\tau\tau}+\omega^2P-\frac{3G\delta}{2}P^2+\frac{G^2}{2}P^3=\delta^2 P_0+\frac{G\delta}{2}P^2_0, \\
\omega^2=\delta^2+h^2-G\delta P_0-\frac{G^2P^2_0}{2}.
\ear
This equation appeared often in the various contexts. In particular, it describes the dynamics of non-degenerate nonlinear dimer (NNLD) related to studies of the phenomenon of self-trapping of the charge and energy carriers between the two sub-units in condensed media \cite{scott,tsir1,kenkbook,kenc,tsir2,tsir3,kapor}. In the present case, the full information on the system dynamics requires the consideration of the dynamics of the EM field in parallel with the temporal evolution of population inversion. To account for MW pulse dynamics, we employ the results from the preceding chapter (\ref{curr}) which yields $I=-\frac{l}{L}\frac{\partial \varphi(x,t)}{\partial x}$, and after the passing to moving frame acquires the form: $I(\tau)= \frac{l}{vL}\frac{\partial \varphi(\tau)}{\partial \tau }$. Finally,  by virtue of equation (\ref{I1}) we got:  \bel{I} I(\tau)=I_0(v)(P_0-P(\tau)),\; I_0(v)=\frac{\hbar g d}{2LC s^2(1-\frac{v^2}{s^22})}.\ee
Apparently, normalized current $i(\tau)=I(\tau)/I_0$ is simply shifted negative of population inversion: $i(\tau)=P_0-P(\tau)$

\section{Solutions}
The character of the solutions of equation (\ref{nlnd}) is determined by parameter 
$\delta$. It may be varied on-demand, providing the means for control of the state of a qubit, and, in the final instance, the pulse propagation. Particularly illustrative case is the resonance $\delta =0$ when  equation (\ref{nlnd}) reduces to that of degenerate nonlinear dimer:
\barl{nldd}
\non && P_{\tau\tau}+\omega^2P+\frac{G^2}{2}P^3=0,\\
&&\omega^2=h^2-\frac{G^2P^2_0}{2}. 
\ear
This equation has been extensively studied regarding charge and energy transfer between the two subunits in complex condensed media \cite{kenc,kenkbook,tsir1,tsir2,tsir3,kapor}. It possesses exact solutions in terms of Jacobi elliptic function according to which transition probability between opposite polarity of persistent current ($CW \rightleftarrows CCW$), depending on the ratio of the nonlinearity parameter ($G(v)$) over the tunneling frequency ($h$), displays the whole scale of different behaviors varying from fast harmonic oscillations to self-trapping \cite{kenc,kenkbook}. The pulse velocity may have a substantial role in the transition between the different dynamical regimes due to quasi-relativistic form $\sim 1/(1-v^2/s^2)$ of the nonlinear parameter (\ref{nlb}), which diverges when $v\rightarrow s$.

We now interrelate these known results with the propagation of the MW pulse within the present setup. For detailed analysis, we now find explicit results for $i(\tau)$ and $P(\tau)$, and in the first step, we evaluate the first integral of eq. (\ref{nldd}):
\bel{first}
P^2_{\tau}=\left(P^2_0-P^2\right)\left[ h^2-\left(\frac{G}{2}\right)^2+\left(\frac{G}{2}\right)^2P^2)\right].
\ee
Introducing the Jacobian modulus ($k=G(v)/2h$) and complimentary modulus ($k'=\sqrt{1-k^2}$), the explicit solutions in terms of Jacobi elliptic functions \cite{ellf} reads:
\barl{ell}
\non h\tau&=&\int_{P/P_0}^{1}\frac{d z}{\sqrt{(1-z^2)(k^2_1+k^2z^2)}}\\
&=&\mathrm{cn}^{-1}(P/P_0|k),\; 0\leqq k \leqq 1,
\ear
Its inversion yields 
\bel{cn}
P(\tau)=P_0\mathrm{cn}(h\tau|k).
\ee

When $k>1$ solutions of eq. (\ref{nldd}) are given through the reciprocal modulus $\mu =1/k$ as follows $\mathrm{cn}(h\tau|k)=\mathrm{dn}(k^{1/2}h\tau, \mu)$ \cite{ellf} yielding 
\bel{dn}
P(\tau)=P_0\mathrm{dn}(k^{1/2}h\tau| \mu).
\ee
Functions $cn$ and $dn$ are periodic with period $4K(k)$ where 
\bel{T}
K(k)=\int_{0}^{\pi/2}\frac{d\theta}{\sqrt{1-k^2\sin^2\theta}}, 
\ee
denotes the elliptic integral of the first kind. Thus, the period of oscillations is $T=K(k)/h$. $k^2=1$ represents the critical point of system dynamics. It displays a sharp transition between two substantially different regimes described by functions- (\ref{cn}) and (\ref{dn}).

\begin{figure}[h]
	\includegraphics[width=8cm]{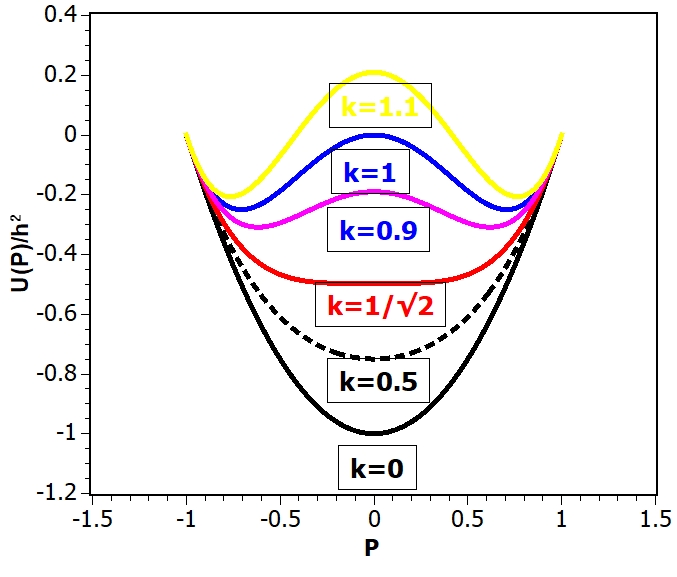}
	\caption{Potential energy of fictitious classical particle vs population inversion.}\label{u}
\end{figure}

\begin{figure}[h]
		\begin{center}
	\includegraphics[width=7cm,height=6cm]{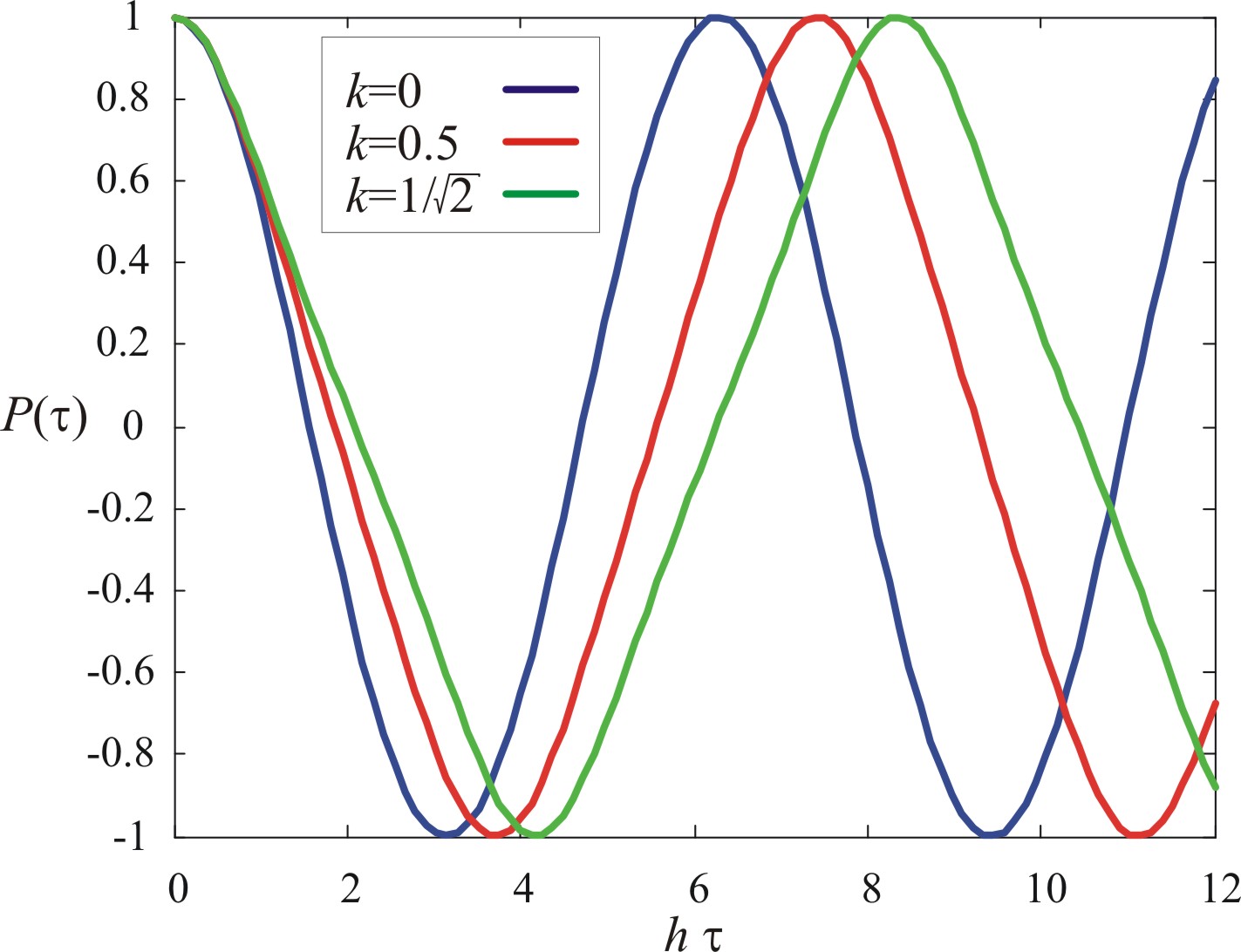}
\vskip 4mm
		\includegraphics[width=7cm,height=6cm]{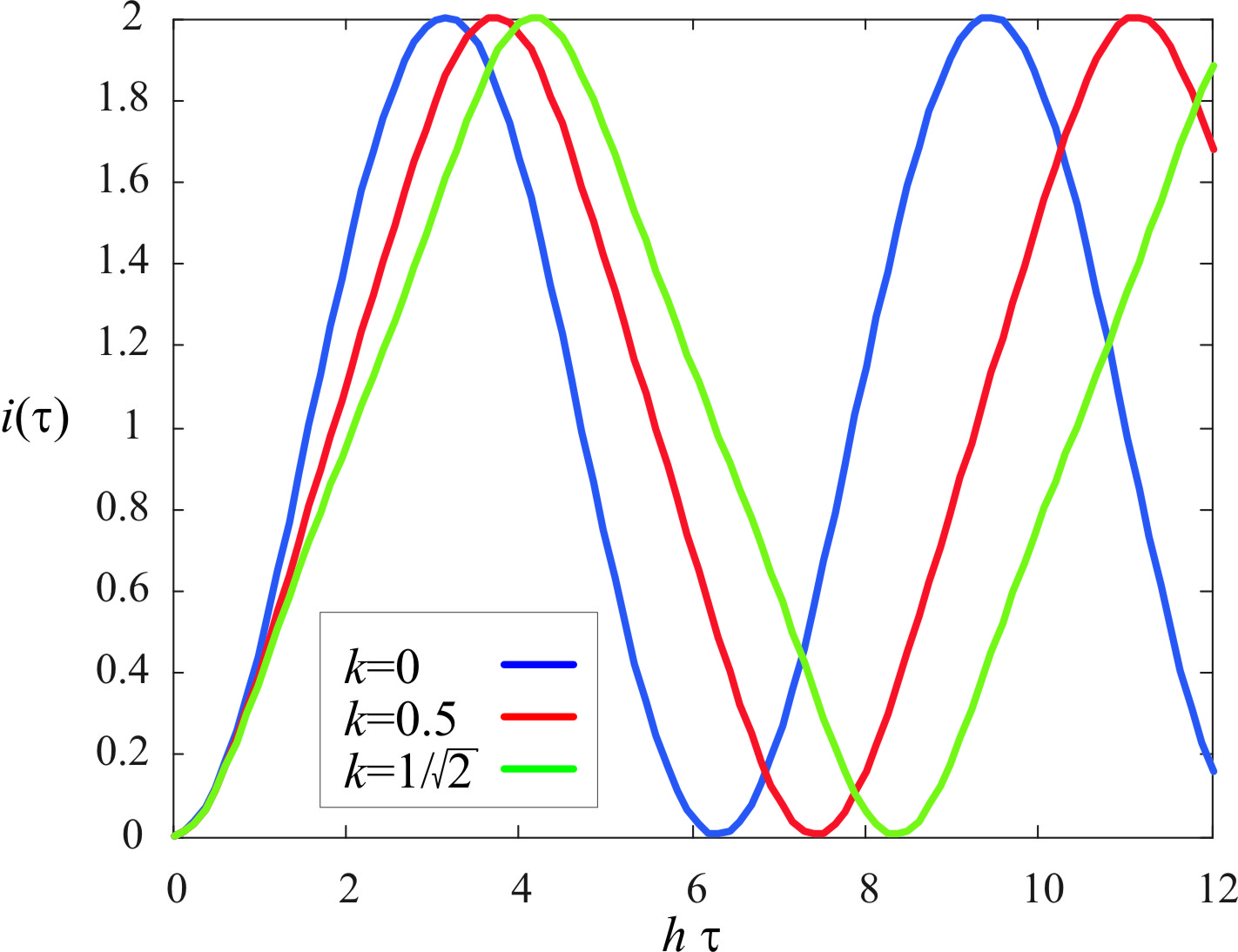}
			\end{center}
	\caption{Graphical illustration of the solutions of equation (\ref{cn}) corresponding to the temporal dynamics of the population inversion (upper panel) and normalized current (lower panel) in the case of weak nonlinearity: $0\leq k \leq 1/\sqrt{2}$, and initial condition $P_0=1$.   }\label{full}
\end{figure}

\begin{figure}[h]
	\begin{center}
\includegraphics[width=7cm,height=6cm]{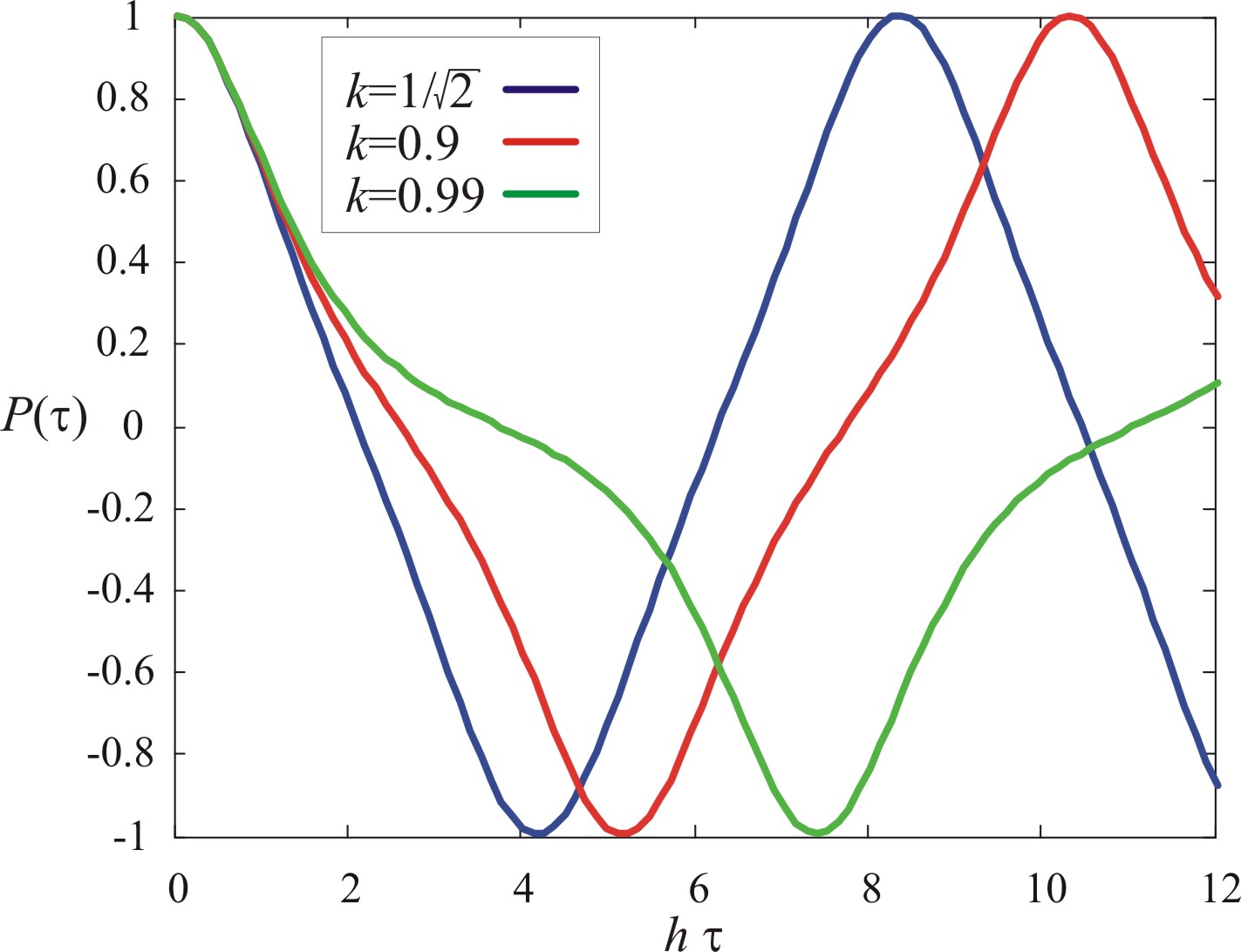}
	\vskip 4mm
		\includegraphics[width=7cm,height=6cm]{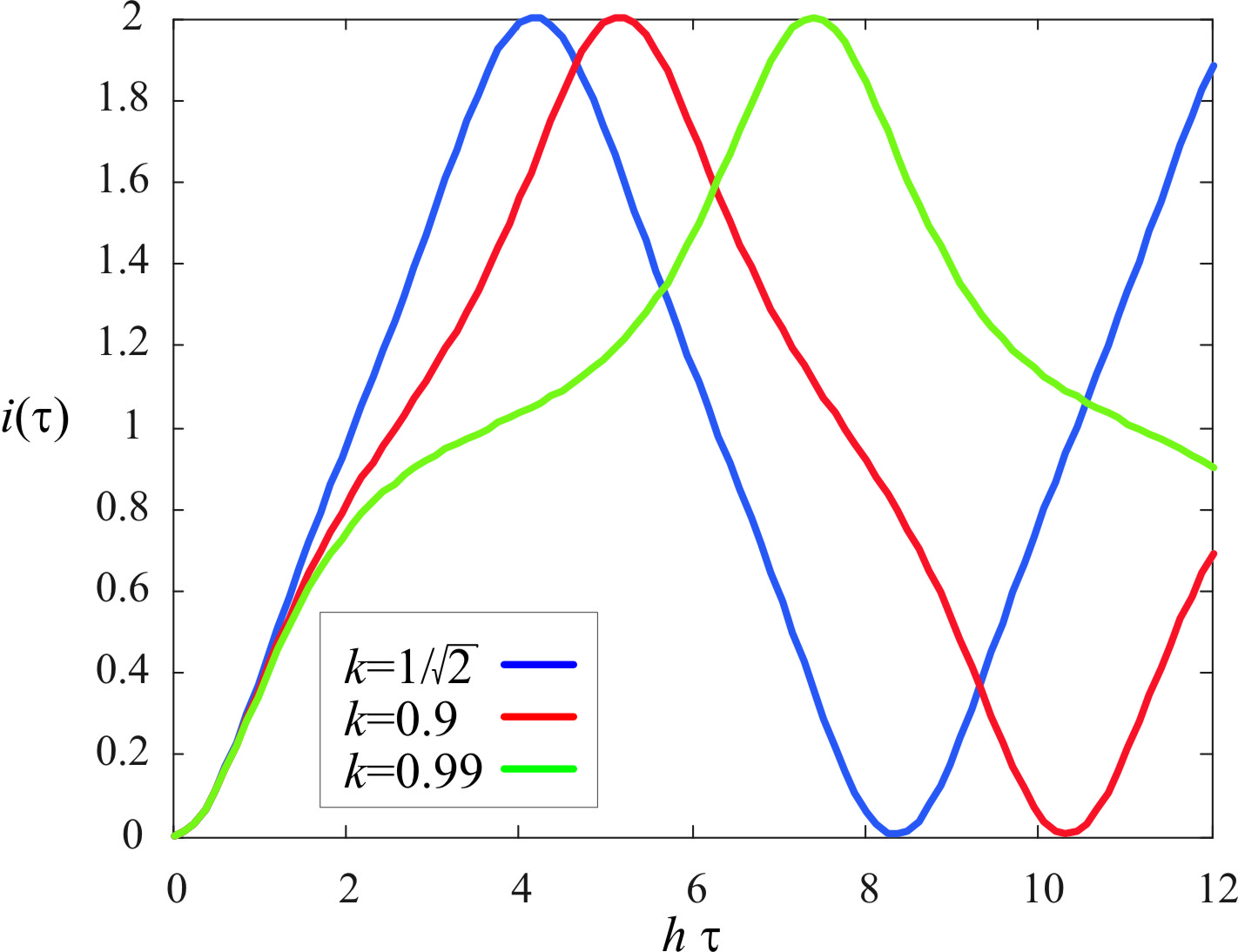}
			\end{center}
	\caption{Visualization of the temporal dynamics of the population inversion (upper panel) and normalized current (lower panel) for intermediate nonlinearity $\sqrt{2}\leq k \leq 1$ and $P_0=1$.  (Equation (\ref{cn}) }\label{full1}
\end{figure}

\begin{figure}[h]
	\begin{center}
	\includegraphics[width=8cm]{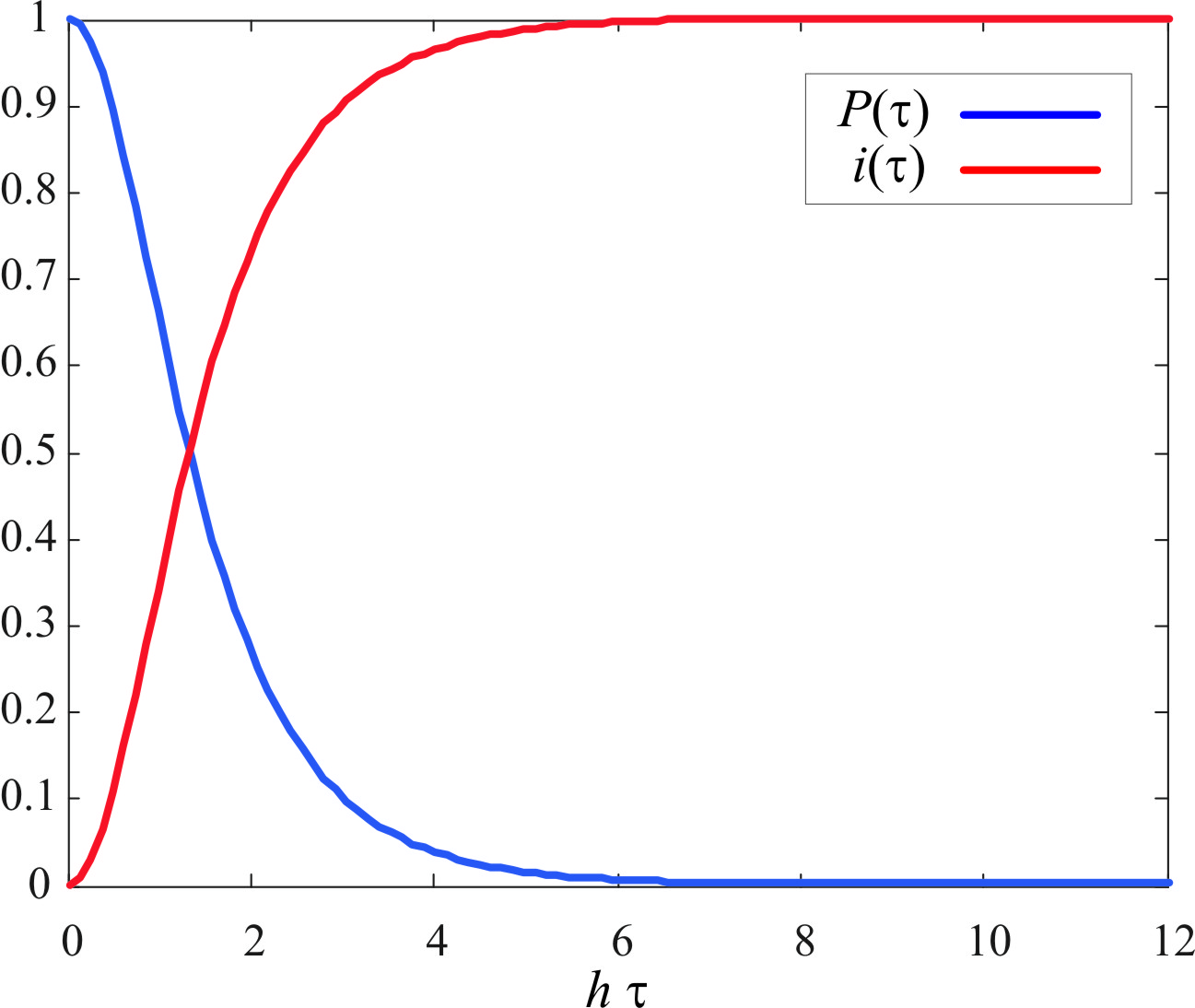}
	\caption{Temporal dynamics of population inversion (blue curve) and normalized current (red curve)  in critical regime (specified by equations (\ref{cn}) and (\ref{dn}) in the limit $k=1$ ).) for qubit subsystem initially prepared in CW state $P_0=1$}\label{tran}
		\end{center}
\end{figure}

\begin{figure}[h]\begin{center}
	\includegraphics[width=7cm,height=6cm]{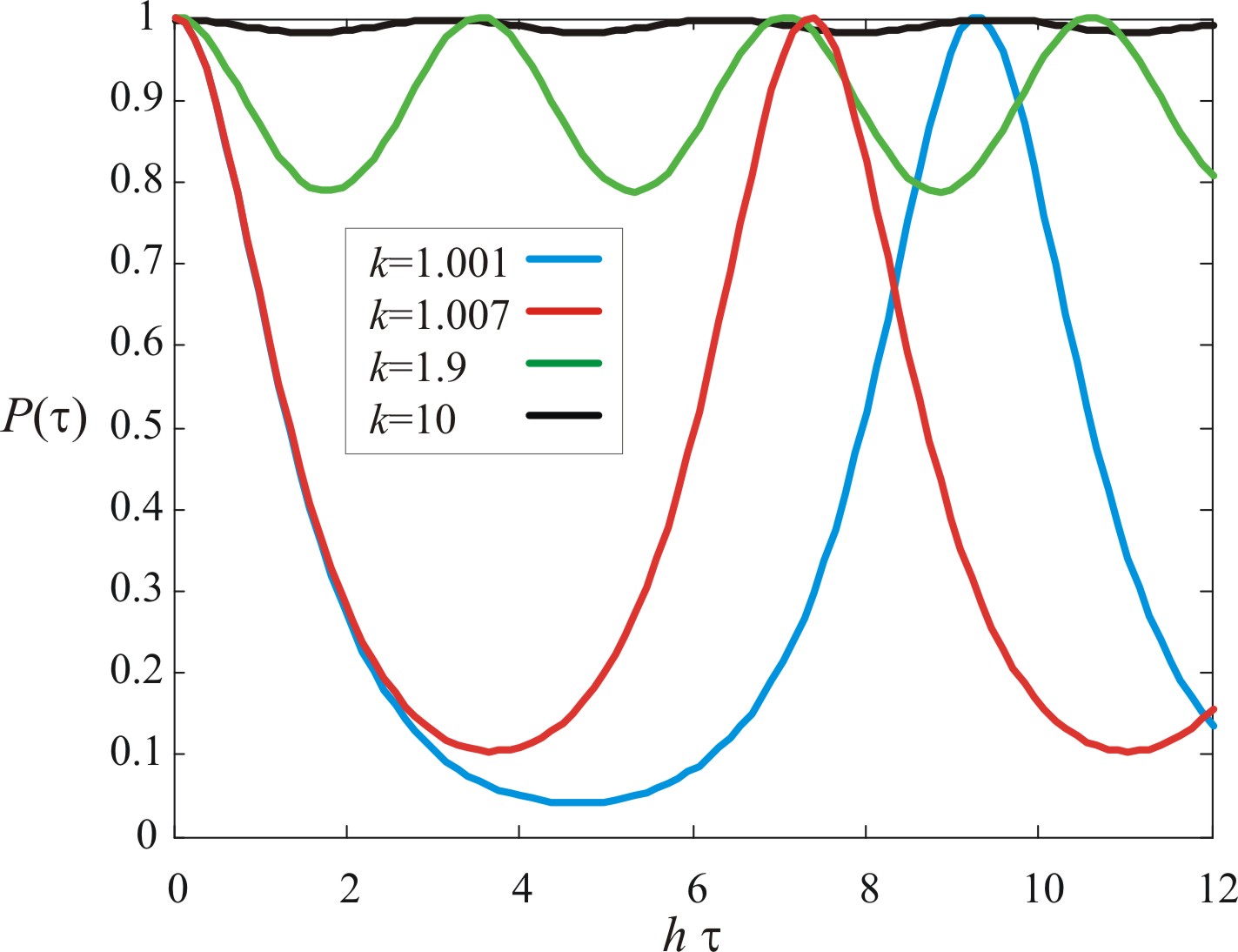}
	\vskip 4mm
		\includegraphics[width=7cm,height=6cm]{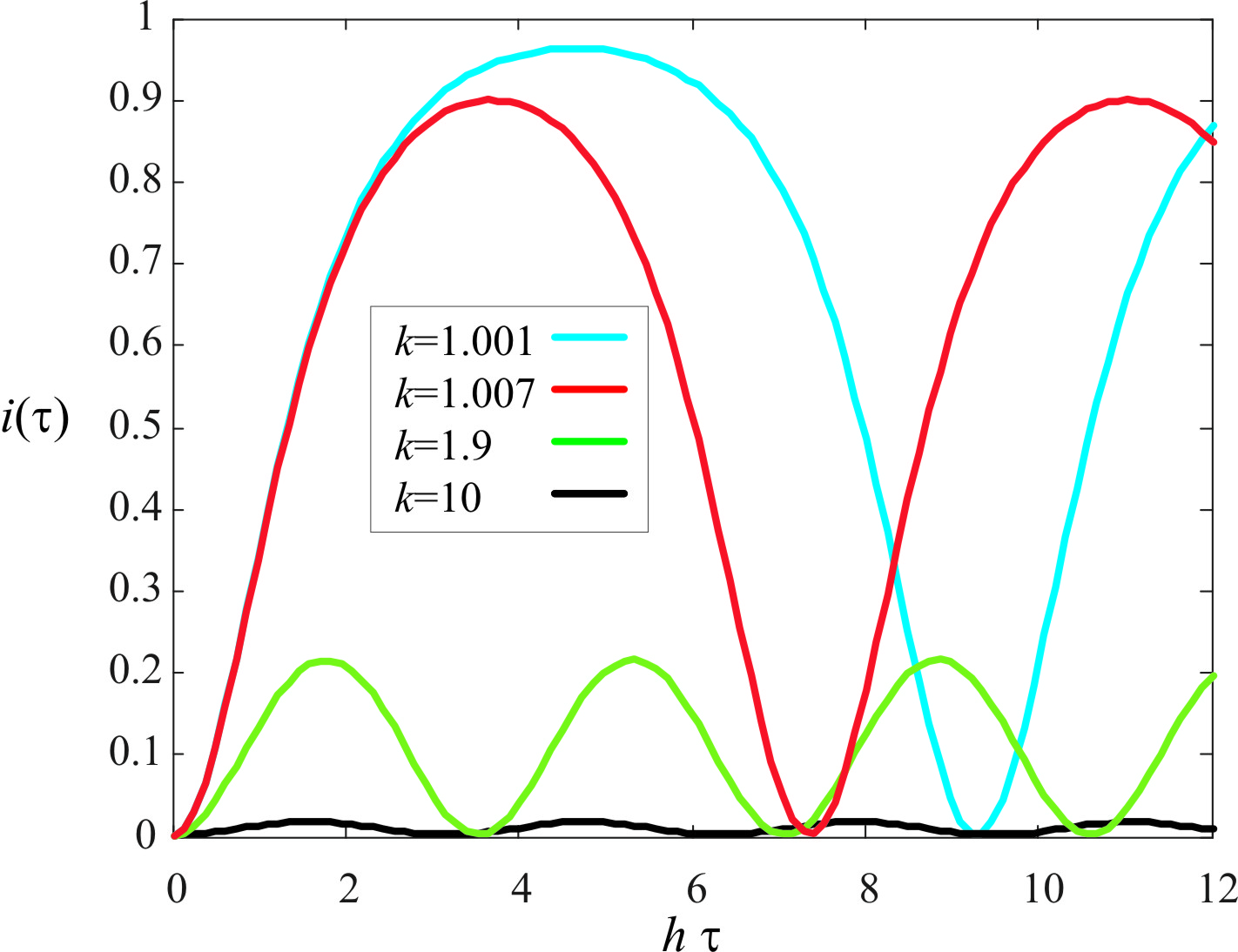}
	\caption{Temporal dynamics of population inversion and current in  nonlinear regime $k>1$ for initial preparation $P_0=1$.}\label{st}	\end{center}
\end{figure}

We now exploit the known analogy with classical mechanics to estimate characteristic values of $k$ for which a particular type of dynamics occurs. To this end, we recall that equation (\ref{first}) may be considered as the total energy of a fictitious classical particle in a double well potential: $E=E_k+U(P)\equiv 0$, where $E_K=P^2_{\tau}$  stands for the "kinetic energy" while 
\bel{pot} U(P)=h^2(P^2-1)[1-k^2+k^2P^2)], \ee 
denotes "potential" energy. We present its explicit form in Fig.(\ref{u}) for a few different values of $k$. 
We visualized our results in figures (\ref{full}-\ref{st}) where we have presented the time evolution of population inversion per particle $P(\tau)$ in parallel with normalized electric current ($I/I_0$) for a few characteristic values of Jacobian modulus. Due to the following symmetry property $P(\tau; P_0=-1)=-P(\tau; P_0=1,$ only the case of qubits initially all polarized clockwise ($P_0=1$) is presented. 

Temporal dynamics of population inversion and normalized current, depending on the particular value of $k$, display three essentially different regimes.

\textit{Full oscillatory reversal of persistent current polarity-$0<k<1$. }

In this case, the population inversion of each FQB displays periodic reversal between CW and CCW states-($P_0 \rightarrow
 - P_0,\; \mathrm{for \; each}\; \tau=(2n-1)T/2, n=1,2,... $). At the same time, normalized current oscillates in antiphase with $P(\tau)$ while its amplitude goes from $0$ to $2$ when all qubits initially are prepared in CW states. Otherwise, it oscillates between $-2$ to $0$. In this interval, there are two subareas in which these oscillations exhibit different features - Fig. (\ref{u}).
The first one corresponds to the interval  $0\leq k \leq 1/\sqrt{2}$ where potential energy has a single minimum at $P=0$ and the form of the parabola whose bottom flattens with the increase of $k$ - Fig.(\ref{u}). Temporal dynamics of both, $P(\tau)$ and $I(\tau)/I_0$, has almost harmonic oscillations whose period ( $T\approx (2\pi/h)(1+1/2\pi)$) enlarges as $k$ increases-Fig. (\ref{full}). In the second subarea ($ 1/\sqrt{2}<k<1$), potential energy attains a double well form, with two degenerate minima at $P_{1,2}=\pm \sqrt{1/(2k^2)-1}$ separated with "potential barrier" at $P=0$. Full oscillations require that the kinetic energy of the "particle" exceeds barrier height, which is fulfilled as far as $k < 1$. Barrier hinders the oscillations between CW and CCW states by increasing their period. That is, both population inversion and current still display oscillatory character. However, as the Jacobian modules increase, the oscillations slow down while their period increases, tending to infinity as the modulo approaches unity-Fig.(\ref{full1}).

\textit{Critical nonlinearity ($k=1$)-stationary regime. Fig.(\ref{tran})}
In this limit, the period of oscillation of population inversion and current tends to infinity, so the blocking of oscillations of both takes place, and the population inversion of each qubit gradually tends to zero. This is a consequence of the gradual decrease in the survival probability of qubit in the state of the initial polarity of the persistent current. At the same time, the transition probability to the state of opposite polarity of persistent current gradually increases. When the equilibrium is reached, the qubit system remains in this mixed state forever -$|\Psi^+(\tau)|^2=|\Psi^-(\tau)|^2$. At the same time, the current attains saturation value at $I/I_0=\pm 1$, and the regime of constant DC is reached. 

\textit{Overcritical $k>1$ nonlinearity - oscillation revival -  and self-trapping transition, partial polarization preservation}.
In this case, the kinetic energy of the fictitious particle is too small to overcome the potential barrier. Thus, it oscillates in a potential well in which it was initially prepared. These oscillations correspond to a partial reversal of the persistent current  polarity. That is, population inversion pulsates between $P_0$ and $0$ according to  $P(\tau)=P_0 dn(k^{1/2}\tau|1/k)$. That is, if the qubit subsystem is initially prepared in a CW (CCW) state, i.e. $|\Psi^{\pm}(\tau_0)|^2=1 $, full polarity reversal to the qubit state of opposite polarity is impossible since  $|\Psi^{+}(\tau)|^2 - |\Psi^{-}(\tau)|^2>0$ ($|\Psi^{+}(\tau)|^2 - |\Psi^{-}(\tau)|^2<0$) for all times. At the same time, $i(\tau)$ attains the character of pulsating DC oscillating in the antiphase with $P(\tau)$. As $k$ rises, the tendency towards self-trapping becomes more pronounced so that population inversion gradually attains the character of very fast small amplitude oscillations in the vicinity of $P_0$ according to $P(\tau)=P_0(1-2/k^2 \sin^2 G\tau) \rightarrow P_0)$, as depicted in Fig.(\ref{st}). In this way, the qubit subsystem remains locked in the initially prepared state while $i(\tau)$ exhibits rapid oscillations in the vicinity of $i(\tau)=0$, which implies the blocking of propagation of the EM pulse.


\section{discussion}
We now discuss the feasibility of the presented results. Note that all our conclusions steam the equation (\ref{nldd}), which holds under the resonance condition-$\delta=0$ or explicitly:
\bel{dissc1} \Delta+\frac{G_0P_0}{1-\frac{v^2}{s^2}}=0.\ee 
The above relation holds both for subluminal ($v<s$) and superluminal ($v>s$) pulses for $\Delta P_0<0$  and  $\Delta P_0>0$ respectively.

The explicit dependence of the nonlinearity parameter on pulse velocity implies the possible self-control of pulse transparency. That means that, for a fixed values of system parameters  ($\Delta, h \; and \; G_0$), each of these regimes may be reached by varying pulse velocity. Nevertheless, the explicit dependence of $k$ on pulse velocity disappears due to above resonance condition (\ref{dissc1}). Thus, $v$ is not free parameter, but it follows that, for each set of system parameters and initial conditions, resonance condition is satisfied for a specific velocity
\begin{equation}\label{v}
	v= s\sqrt{1+\dfrac{G_0P_0}{\Delta}}.  
	\end{equation}
Once when this velocity is achieved, the above predicted regimes of dynamics of pulse dynamics may be controlled by just two parameters as can be seen from the following relation 
	\bel{k} k =- \frac{\Delta P_0}{2 h},
\ee
which follows after the substitution of (\ref{v}) into the expression for Jacobi elliptic module.

\section{Concluding remarks}
In this paper, we have investigated the propagation of a MW pulse through a flux qubit-based QMM. The transmission of the pulse through the QMM is accompanied by the reversal of the persistent current polarity in the qubits. The system exhibits three distinct dynamical regimes: (i) quasi-harmonic oscillations of population inversion and the associated current in the weak nonlinearity regime ($k<1$), (ii) stationary transmission at $k=1$, and (iii) complete suppression of population inversion and current in the strong nonlinearity regime ($k>1$).

The present analysis has been carried on in idealized conditions neglecting the relaxation effects coming from coupling with the environment that would lead to energy loss in qubit subsystem and pulse decay in final instance. This could be realized for very short pulses whose duration time is small in comparison with longitudinal and transverse relaxation time characterizing,
respectively, the decay of “atoms” from excited to ground state and pure dephasing. The validity of adiabatic approximation is also related with this kind of relaxation processes. Namely, equation (\ref{I1}), is approximate in the sense that we took only the particular solution while neglecting the homogeneous part. It corresponds to a fluctuations in the EM field and play the role thermal bath. Its accounting for leads to a system of damped stochastic nonlinear Bloch equations similar to that studied in \cite{kapor}. Thus, the validity of adiabatic approximation is provided that relaxation time highly exceeds pulse duration.

Further idealization concerns the neglecting of the structural non-homogeneities of FQB energy levels due to non–uniform sizes of JJs stemming from fabrication conditions of FQBs. Their influence is similar to the effects of inhomogeneous broadening of atom levels in quantum optics \cite{alleb} where it play even constructive role in the emergence of phenomena such as self-induced transparency and photon echo \cite{alleb}.

We finally emphasize that the predicted effects rely on long-distance photon interaction with flux qubits defined through the specific form of interaction Hamiltonian (\ref{int}), and  are applicable only for QMM based on flux qubits. Specifically, the evolution equation for the flux variable, in the continuum approximation, simplifies to the form specified by the third equation in system (\ref{nlbloch}) which can be exactly integrated in adiabatic approximation. In the case of different type of qubits interaction Hamiltonian would have quite different form. 

\acknowledgements{This research was funded by Ministry of Science, Technological Development and Innovations of the Republic of Serbia (Contract No. 451-03-136/2025-03/ 200017)}


\begin{thebibliography}{99}
	\bibitem{qub0}
	Uri Vool
	and Michel Devoret
	Introduction to quantum electromagnetic circuits,
	Int. J. Circ. Theor. Appl. (2017)
	\bibitem{qub01}
	Girvin SM. 2014. In Quantum Machines: Measurement and Control of Engineered Quantum Systems, Proceed-
	ings of the 2011 Les Houches Summer School on Quantum Machines, ed. M Devoret, B Huard, R Schoelkopf,
	LF Cugliandolo, pp. 113–256. Oxford, UK: Oxford Univ. Press
	\bibitem{qub1}
		J. M. Martinis, M. H. Devoret, and J. Clarke, Phys. Rev. Lett.,
	Energy-Level Quantization in the Zero-Voltage State of a Current-Biased Josephson Junction
	Phys. Rev. Lett. 55, 1543 (1985),
	
\bibitem{qub2}
Nakamura Y, et al, 1999 Coherent control of macroscopic quant states in a single-Cooper-pair
box, Nature 398 786-8

\bibitem{qub3}
van der Wal C.H, ter Har A.C.J et al, Quantum superposition of macroscopic persistent-current states, 2000 Science 290 773-7

\bibitem{qub4}
Chiorescu I, Nakamura Y, Hermans C.J.P.M and Mooj J E,
Coherent Quantum dynamicss of a superconducting flux qubit, 2003, Science 299 1869-71,

\bibitem{lind}
T. Lindstr\"om, et al.
Circuit QED with a flux qubit strongly coupled to a coplanar transmission line resonator,
Supercond. Sci. Technol. \textbf{20}, 814-821, 2007.
\bibitem{qub5}
A. N. Omelyanchouk, et al. Quantum behavior of a flux qubit coupled ta resonator,
Low temperature physics 36, 893, 2010,

\bibitem{qub6}
Jonathan R. Friedman, Vijay Patel, W. Chen, S. K. Tolpygo \& J. E. Lukens, Quantum superposition of
distinct macroscopic states,
NATURE, VOL 406, 43, 2000,

\bibitem{qub7}
Yannick Schön, Jan Nicolas Voss, Micha Wildermuth, Andre Schneider, Sebastian T. Skacel, Martin P. Weides, Jared H. Cole, Hannes Rotzinger1, and Alexey V. Ustinov,
Rabi oscillations in a superconducting nanowire circuit
 npj Quantum Mater. 5, 18 (2020)
 \bibitem{qub8}
T. Katsuzawa, et al.
Coherent control of a flux qubit by phase-shifted resonant microwave pulses,
App.Phys.Lett 87, 073501, 2005 

\bibitem{cqed1}Andre Blais, Arne L. Grimsmo, S. M. Girvin, Andreas Wallraff,
Circuit quantum electrodynamics,
Rev. Mod. Phys. {\bf 93}, 025005 (2021).

	\bibitem{cqed2}
		P. Krantz, M. Kjaergaard, F. Yan, T. P. Orlando, S. Gustavsson,1 and W. D. Oliver
	A quantum engineer's guide to superconducting
	qubits
	Appl. Phys. Rev. 6, 021318 (2019).
	
	\bibitem{cqed3} 
	Michael Reitz, Christian Sommer and Claudiu Genes 
	Cooperative Quantum Phenomena in Light-Matter Platforms,
	PRX Quantum {\bf 3}, 010201 (2022).
	
	\bibitem{cqed4} M. Mirhosseini, E. Kim, V. S. Ferreira, M. Kalaee, A. Sipahigil, A. J.
	Keller, and O. Painter, “Superconducting metamaterials for waveguide
	quantum electrodynamics,” Nat. Commun 9, 3706 (2018).
	
	\bibitem{cqed5}
	Gu, A. F. Kockum, A. Miranowicz, Y.-x. Liu, and F.
	Nori, Microwave photonics with superconducting quantum
	circuits, Phys.Rept. 718-719, p 1, 2017,
	10.1109/TMTT.2021.3105431
	
	\bibitem{qmm1}
	A. L. Rakhmanov, A. M. Zagoskin, S. Savel'ev, and F. Nori, Quantum metamaterials: Electromagnetic waves in a Josephson qubit line, Phys. Rev. B {\bf 77}, 144507 (2008).
		\bibitem{qmm11}
	Zagoskin et al. Quantum metamaterials: Electromagnetic waves in a Josephson qubit line
	Phys. Status Solidi B 246, No. 5, 955–960 (2009) / DOI 10.1002/pssb.200881568

	\bibitem{qmm2}
	A. M. Zagoskin,\textit{ Quantum Engineering: Theory and Design of Quantum Coherent Structures} (Cambridge: Cambridge University Press) pp. 1--346 (2011).
		
	\bibitem{asai15}
	H. Asai, S. Savel’ev, S. Kawabata, A.M. Zagoskin,
	Effects of lasing in a one-dimensional quantum metamaterial, 
	Phys.Rev. B 91, 134513 (2015)
	
	\bibitem{asai18}
	Hidehiro Asai, Shiro Kawabata, Sergey E. Savel’ev, and Alexandre M. Zagoskin,	
	Quasi-superradiant soliton state of matter in quantum
	metamaterials,
	Eur. Phys. J. B (2018) 91: 30
	
		\bibitem{qmm}
	J. Q. Quach, C. Su, A. M. Martin, A. D. Greentree, and L. C. L. Hollenberg, “Reconfigurable quantum metamaterials,” Opt. Express 19,
	11018–11033 (2011).
	
	\bibitem{qmm10} 
	S I Mukhin and M V Fistul
	
	Generation of non-classical photon
	states in superconducting quantum
	metamaterials
	
	Supercond. Sci. Technol. 26 (2013) 084003 (7pp) doi:10.1088/0953-2048/26/8/084003
	
	\bibitem{qmm3}
	SOLOMON URIRI, YASEERA ISMAIL  AND FRANCESCO PETRUCCIONE
	Quantum Metamaterials: Applications in quantum information science.2020, arXiv: 2006.03757v1
	\bibitem{qmm4}
	
	P. Macha, G. Oelsner, J. M. Reiner, M. Marthaler, S. Andr{\'e}, G. Sch{\"o}n, U. H{\"u}bner, H. G. Meyer, E. Il'ichev, and A. V. Ustinov, Implementation of a quantum metamaterial using superconducting qubits, Nat. Commun. {\bf 5}, 5146 (2014).
	
	\bibitem{qmm5}
	P. Jung, A. V. Ustinov, and S. M. Anlage, Progress in superconducting metamaterials, Supercond. Sci. Tech. {\bf 27}, 073001 (2014).
	
	\bibitem{qmm6} 
	N. Lazarides and G. P. Tsironis, Superconducting metamaterials, Phys. Rep. {\bf 752}, 1 (2018).
	
	\bibitem{qi1}
	M. H. Devoret and R. J. Schoelkopf, Superconducting circuits for quantum information: An outlook, Science {\bf 339}, 1169 (2013).
	
	\bibitem{qi2}
	S. Schmidt and J. Koch, Circuit QED lattices: Towards quantum simulation with superconducting circuits, Ann. Phys. (Berlin) {\bf 525}, 395 (2013).
	
	\bibitem{qi3}
	I. M. Georgescu, S. Ashhab, and F. Nori, Quantum simulation, Rev. Mod. Phys. {\bf 86}, 153 (2014).
	
	\bibitem{qi4}
	J. C. Bardin, D. Sank, O. Naaman, and E. Jeffrey,  Quantum computing: An introduction for microwave engineers, IEEE Microw. Mag., 21, no 8, 24–44, 2022.
	
\bibitem{eit1}
Junling Long,  H. S. Ku, Xian Wu,  Xiu Gu, Russell E. Lake, Mustafa Bal, Yu-xi Liu, and David P. Pappas,

Electromagnetically Induced Transparency in Circuit Quantum Electrodynamics with Nested Polariton States,

Phys. Rev. Lett.   120, 083602, (2018).

\bibitem{eit2}
Lan Zhou, Jing Lu, C. P. Sun,
Coherent control of photon transmission: Slowing light in a coupled resonator waveguide doped
with $\lambda$ atoms,
Phys.Rev. A 76, 012313 2007
\bibitem{sit1}Z. Ivić, N. Lazarides, and G. Tsironis, “Light manipulation by quantum
metamaterials,” Contemp. Mater. 2, 186–189 (2014).

\bibitem{sit2} 
Z. Ivi\'c, N. Lazarides, and G. Tsironis Qubit lattice coherence induced by electromagnetic pulses in superconducting metamaterials, Sci. Rep. {\bf 6}, 29374 (2016).

\bibitem{sit3}
Z. Ivi\'c, D. \v Cevizovi\'c, \v Z. Pr\v zulj, N.Lazarides and G.P.Tsironis,
Dispersive properties of self--induced transparency in two-level media, Chaos, Solitons and Fractals {\bf 143}, 110611 (2021).

\bibitem{sit4}
Zoran Ivi\'c, Nikos Lazarides and G.P.Tsironis,
Self-induced transparency in a flux-qubit chain,
Chaos, Solitons \& Fractals {\bf 1}, 100003 (2019).
\bibitem{slowqmm}
Jan David Brehm, Richard Gebauer, Alexander Stehli, Alexander N. Poddubny, Oliver Sander, Hannes Rotzinger, and Alexey V. Ustinov, Slowing down light in a qubit metamaterial, Appl. Phys. Lett. 121, 204001, (2022)

	\bibitem{fqb1}
Mooij J E, Orlando T P, Levitov L, Tian L,
van der Wal C H and Lloyd S 1999 Josephson
persistent-current qubit Science 285 1036
\bibitem{fqb2}
P. Orlando, J. E. Mooij, L. Tian, C. H. van der Wal, L. S. Levitov, S.
Lloyd, and J. J. Mazo, Phys. Rev. B 60, 15399, 1999.
\bibitem{pozar}
Pozar, David M.
Microwave engineering, Wiley, New York, 3rd ed. 1990.

\bibitem{roth}T. E. Roth and S. T. Elkin, "Maxwell-Schrödinger Modeling of a Superconducting Qubit Coupled to a Transmission Line Network," in IEEE Journal on Multiscale and Multiphysics Computational Techniques, vol. 9, pp. 61-74, 2024, doi: 10.1109/JMMCT.2024.3349433.

\bibitem{ahmad}
R. K. Bulough and F. Ahmad,
Exact Solutions of the Self-Induced Transparency Equations,
Phys. Rev. Lett. 27, 330, (1971),
\bibitem{kaplan}
A.E. Kaplan and P. L. Shkolnikov,
Electromagnetic "Bubbles" and Shock waves, Nonoscillating EM Solitons
Phys. Rev. Lett. 75, 2316, (1995),
\bibitem{magsol}
C. Calero, E. M. Chudnovsky, and D. A. Garanin,
Magneto-elastic waves in crystals of magnetic molecules,
Phys.Rev. B 76, 094419, 2007.
\bibitem{polaron}
A. S. Davydov,
Solitons in Molecular Systems

Springer-Science+Business Media, B.Y.


\bibitem{scott}
A.C. Scott, Phys.Scr. 42 (1990) 14,
\bibitem{tsir1}
Tsironis G (1993) Dynamical domains of a nondegenerate nonlinear dimer. Physics
Letters A 173(4-5):381–385
\bibitem{kenkbook}
V. M. N. Kenkre, Interplay of Quantum Mechanics and Nonlinearity, 1st ed. (Springer, 2022).
\bibitem{kenc}
Kenkre VM, Campbell DK (1986) Self-trapping on a dimer: time-dependent solutions
of a discrete nonlinear Schrödinger equation. Physical Review B 34(7):4959
\bibitem{tsir2}
Tsironis GP, Kenkre VM (1988) Initial condition effects in the evolution of a nonlinear
dimer. Physics Letters A 127(4):209–212
\bibitem{tsir3}
Tsironis GP, Kenkre VM, Finley D (1988) Effects of dissipation on nonlinearity in
transport: Evolution and integrability properties in a molecular dimer. Physical
Review A 37(11):4474
\bibitem{kapor}
D. Kosti\'c, Z. Ivi\'c, D. Kapor, M. Lali\'c and A. Tan\v ci\'c,
Interimpurity transfer in condensed media: Breakdown of coherent tunneling
and conditions for the creation of localized states
Phys. Rev. B50, 13315, (1994).

\bibitem{ellf}
Derek F. Lawden, Elliptic Functions and Applications,
1989 by Springer-Verlag New York Inc.

\bibitem{alleb}
L. Allen, J. Eberly, \textit{Optical resonance and two-level atoms}, (Wiley, New York, 1975).

\end{thebibliography}

\end{document}